\documentclass[paper=A4,11pt,DIV=12,headings=big]{scrartcl}
\pdfoutput=1

\usepackage{amsfonts}
\usepackage{amsmath,bm}
\usepackage{amssymb}
\usepackage[normal]{caption}

\usepackage{cite}
\usepackage{enumerate}
\usepackage[a4paper, top=1.2in, bottom=1.2in, left=1.1in, right=1.1in]{geometry}
\usepackage[multiple]{footmisc}
\usepackage{graphicx}
\graphicspath{{./}{./figures/}}
\usepackage{slashed}
\usepackage{xcolor}

\allowdisplaybreaks
\addtokomafont{disposition}{\rmfamily\boldmath}

\def\l{\ensuremath{\left}}
\def\r{\ensuremath{\right}}

\newcommand*\diff{\mathop{}\!\mathrm{d}}

\newcommand{\I}{\ensuremath{i\mkern1mu}}
\newcommand{\E}{\ensuremath{e\mkern1mu}}

\newcommand{\mc}{\mathcal}
\newcommand{\nno}{\nonumber}
\newcommand{\fr}{\frac}

\begin{document}

\begin{titlepage}
\def\thefootnote{\fnsymbol{footnote}}

\begin{flushright}
  IU-HET-610 \\
  KIAS-P16005
\end{flushright}

\begin{centering}
\vspace{0.5cm}
{\sffamily \bfseries \LARGE \boldmath
Secluded singlet fermionic dark matter \\[0.2cm]
driven by the Fermi gamma-ray excess}

\bigskip

\begin{center}
  {\normalsize \sffamily \bfseries
    Yeong~Gyun~Kim$^{a,b,}$\footnote{ygkim@gnue.ac.kr},
    Kang~Young~Lee$^{c,}$\footnote{kylee.phys@gnu.ac.kr},
    Chan~Beom~Park$^{b,}$\footnote{cbpark@kias.re.kr},
    Seodong~Shin$^{d,}$\footnote{shinseod@indiana.edu}}\\[0.5cm]
  \small
  $^a${\em Department of Science Education, Gwangju National
    University of Education,\newline Gwangju 61204, Korea}\\[0.1cm]
  $^b${\em School of Physics, Korea Institute for Advanced Study,
    Seoul 02455, Korea}\\[0.1cm]
  $^c${\em Department of Physics Education \& Research Institute of
    Natural Science,\newline
    Gyeongsang National University, Jinju 52828, Korea}\\[0.1cm]
  $^d${\em Physics Department, Indiana University, Bloomington, IN
    47405, USA}
\end{center}
\end{centering}

\medskip

\begin{abstract}
  \noindent
 We examine the possibility that the dark matter (DM) interpretation of the GeV scale Fermi gamma-ray excess at the Galactic Center can be realized in a specific framework - secluded singlet fermionic dark matter model with small mixing between the dark and Standard Model sector. Within this framework it is shown that the DM annihilation into bottom-quark pair, Higgs pair, and new scalar pair, shown to give good fits to the Fermi gamma-ray data in various model independent studies, can be successfully reproduced in our model. Moreover unavoidable constraints from the antiproton ratio by the PAMELA and AMS-02, the gamma-ray emission from the dwarf spheroidal galaxies by the Fermi-LAT, and the Higgs measurements by the LHC are also considered. Then we found our best-fit parameters for the Fermi gamma-ray excess without conflicting other experimental and cosmological constraints if uncertainties on the DM density profile of the Milky Way Galaxy are taken into account. Successfully surviving parameters are benchmark points for future study on the collider signals.
\end{abstract}

\vspace{0.5cm}

\end{titlepage}

\renewcommand{\thefootnote}{\arabic{footnote}}
\setcounter{footnote}{0}

\setcounter{tocdepth}{2}
\noindent \rule{\textwidth}{0.3pt}\vspace{-0.4cm}\tableofcontents
\noindent \rule{\textwidth}{0.3pt}

\section{Introduction}
\label{sec:intro}

The existence of non-baryonic dark matter (DM) in the universe has been supported by a lot of solid evidences observing from its gravitational interactions. On the other hand its particle property is still in a mystery. Among all the candidates Weakly Interacting Massive Particle (WIMP) is the most popular one because of its natural mass and interaction ranges to give the right amount to account for the matter density observed today.
Experimental efforts to detect WIMPs are ongoing in direct and indirect searches as well as collider experiments.
Direct detection experiments are designed to observe the elastic scattering of WIMPs on the target nuclei through nuclear recoils.
On the other hand, indirect detection experiments search for products of the WIMP annihilation or decay processes such as gamma-rays, neutrinos, and charged cosmic rays. Among those products, gamma-rays are often considered as the golden channel for the indirect detection of the DM because we can easily detect them and identify in which part of the universe they came from.

It is intriguing that several independent collaborations have reported a broad excess of the gamma-ray (at energies around few GeV) from the Galactic Center (GC) above the expected astrophysical emission through the analyses of the data accumulated by
the Fermi Large Area Telescope (LAT)~\cite{Goodenough,Hooper:2011ti,Abazajian:2012pn,Hooper:2013rwa,Gordon:2013vta,Abazajian:2014fta,Daylan:2014rsa,Calore:2014xka,Calore:2014nla}, which is confirmed later by the experimental group \cite{TheFermi-LAT:2015kwa}.
The excess might be explained by (unidentified) astrophysical sources such as millisecond pulsars~\cite{Gordon:2013vta,Abazajian:2014fta,msp} which can be an important part of unresolved point sources fitting the observed data \cite{Bartels:2015aea,Lee:2015fea}. However the DM annihilation still remains as a most viable possibility to account for it \cite{TheFermi-LAT:2015kwa}. Many collaborations have investigated such a possibility in a model independent way by classifying various scenarios of the DM on the basis of the final states of the annihilation process, and in turn quantitatively obtaining the scales of the mass and the annihilation cross section of the DM that can fit the gamma-ray excess. As a result, it has been shown that the DM annihilations into a pair of $b$ quarks~\cite{Daylan:2014rsa}, leptons~\cite{Lacroix:2014eea}, the Higgs bosons~\cite{Agrawal:2014oha,Calore:2014nla} in the Standard Model (SM), or new particles which subsequently decay to $b \bar b$ pairs~\cite{Abdullah:2014lla,Cline:2015qha,Elor:2015tva,Rajaraman:2015xka,Ko:2015ioa,Dutta:2015ysa} can give good fits to the data with proper choices of DM mass and the annihilation cross section.\footnote{See also \cite{Kim:2015usa} for the contribution by a diphoton production in a dark sector cascade decay with two or more DM candidates whose mass gaps are non-negligible.}

The model independent studies typically consider the DM mass and the annihilation cross section as free parameters to fit the gamma-ray data for the given annihilation process. In this case, it is obvious that the impact on other experimental and cosmological constraints is subtle. Hence, in the end, one has to consider a specific model that includes a plausible DM candidate and study various other theoretical and experimental bounds in connection with the gamma-ray excess. Preceding model dependent analyses exist with the main annihilation channel for the GeV scale excess; $b$-quark pair~\cite{Alvares:2012qv,Okada:2013bna,Modak:2013jya,Alves:2014yha,Ipek:2014gua,Basak:2014sza,Wang:2014elb,Ghorbani:2014qpa,Cao:2014efa,Ghorbani:2014gka,Ghorbani:2015baa,Duerr:2015bea}, lepton pair~\cite{Kyae:2013qna,Kim:2015fpa}, and new particles~\cite{Boehm:2014bia, Berlin:2014pya,Cerdeno:2015ega,Cao:2015loa}.

In this paper, we examine if the DM interpretation for the gamma-ray excess
can be realized in a model with a singlet fermionic dark matter (SFDM) which is originally proposed in~\cite{Kim:2008pp}. In this model, DM interacts with the SM sector only by the mixing between the SM Higgs and a singlet scalar. Particularly we specify the scenario where such a mixing is quite suppressed to avoid the recent bounds from the Higgs measurements at the Large Hadron Collider (LHC) and various direct detection results of WIMP, while keeping the relic density as observed.\footnote{Additional annihilation channels exist at the freeze-out other than the $s$-channel scalar exchanges.} Actually this secluded SFDM scenario was previously suggested in~\cite{Kim:2009ke} to provide a viable light WIMP setup.\footnote{See also \cite{lightwimp} for the summary of the related issue.} Here we slightly modify the secluded SFDM scenario by adding a pseudoscalar interaction in the dark sector to easily explain the Fermi gamma-ray excess.

Interestingly, following the analysis in~\cite{Calore:2014nla}, we could find the parameters giving good fits to the excess in several DM annihilation channels; $b$ quark pair, Higgs pair, and new scalar pair (without fixing the decay mode of it by hand), depending on the mass hierarchies of particles and couplings. On top of these we further consider unavoidable constraints from the antiproton ratio by the PAMELA~\cite{Adriani:2010rc} and AMS-02~\cite{AMS02,Giesen:2015ufa}, the gamma-ray emission from the dwarf spheroidal galaxies by the Fermi-LAT~\cite{Ackermann:2015zua}, and the Higgs measurements by the LHC~\cite{atlascms}. These bounds are quite strong so the annihilation channel to $b$ quark pair remains viable only around the resonance region and after introducing a mixture of the scalar and pseudoscalar interaction in the dark sector. The surviving parameters in all the channels will be our benchmark points for future study on the collider signals.

The rest of the paper is organized as follows.
The description for the SFDM model is given in Sec.~\ref{sec:model}. Then
in Sec.~\ref{sec:gamma_excess}, we calculate the photon energy
spectrum from the SFDM annihilations, and in turn perform the fit to the
gamma-ray data by considering other experimental and cosmological
constraints. Sec.~\ref{sec:concl} is devoted to conclusions.

\section{Singlet fermionic dark matter}
\label{sec:model}

We introduce a real scalar field $S$ and a Dirac fermion field
$\psi$, which transform as the singlet under the SM gauge group. In addition to the SM Lagrangian, the dark sector Lagrangian
with the renormalizable interactions is given by
\begin{align}
  \mc{L}^{\rm dark} = \bar\psi (\I \slashed{\partial}
   - m_{\psi_0}) \psi + \fr{1}{2} \partial_\mu S \partial^\mu S  - g_S
  (\cos\theta \,\bar\psi \psi + \sin\theta \,\bar\psi \I \gamma^5 \psi) S
  - V_S (S, \, H),
  \label{eq:lagrangian}
\end{align}
where
\begin{align}
  V_S (S, \, H) = \fr{1}{2} m_0^2 S^2 + \lambda_1 H^\dagger H S +
  \lambda_2 H^\dagger H S^2 + \fr{\lambda_3}{3!} S^3 +
  \fr{\lambda_4}{4!} S^4 .
\end{align}
The interactions of the singlet sector to the SM sector arise
only through the Higgs portal $H^\dagger H$ as given above.
Note that we extend the model proposed in
\cite{Kim:2008pp,Kim:2009ke,Kim:2006af} by including the pseudoscalar interaction in the singlet sector.
The inclusion of the pseudoscalar interaction is helpful for the analysis on the gamma-ray excess since it can conveniently fit the excess while satisfying the other constraints as explained in more detail in the next section.\footnote{See also Refs.~\cite{Fedderke:2014wda,LopezHonorez:2012kv}
for the related study.}

Together with the SM Higgs potential,
\begin{align}
  V_\mathrm{SM} = -\mu^2 H^\dagger H + \lambda_0 (H^\dagger H)^2,
\end{align}
the Higgs boson acquires a vacuum expectation value (VEV) after
electroweak symmetry breaking (EWSB), and it can be written in the unitary
gauge as
\begin{align}
  H = \fr{1}{\sqrt{2}}
  \begin{pmatrix}
    0 \\ v_h + h
  \end{pmatrix}
\end{align}
with $v_h = (\sqrt{2}G_F)^{-1/2} \simeq 246$~GeV. The singlet scalar
field also develops a non-zero VEV $v_s$, so we expand the singlet
scalar field around the VEV as $S = v_s + s$.
The mass parameters $\mu^2$ and $m_0^2$ can be eliminated by using the
minimization conditions of the full potential of the scalar fields,
$V_S + V_\mathrm{SM}$. The relations are given as follows.
\begin{align}
  \mu^2 &= \lambda_0 v_h^2 + (\lambda_1 + \lambda_2 v_s) v_s ,\nno\\
  m_0^2 &= - \l( \fr{\lambda_1}{2 v_s} + \lambda_2 \r) v_h^2
          - \l( \fr{\lambda_3}{2 v_s} + \fr{\lambda_4}{6} \r) v_s^2 .
\end{align}
In this setup, the mass term for the scalar fields $\Phi^\mathsf{T} = (h, \, s)$
is
\begin{align}
  \mc{L}_\mathrm{mass} = - \fr{1}{2} \Phi^\mathsf{T} \mc{M}_\Phi^2 \Phi
  = -\fr{1}{2}
  \begin{pmatrix}
    h & s
  \end{pmatrix}
  \begin{pmatrix}
    \mu_h^2 & \mu_{hs}^2\\
    \mu_{hs}^2 & \mu_s^2
  \end{pmatrix}
  \begin{pmatrix}
    h \\ s
  \end{pmatrix}
  \label{eq:higgs_mass_matrix}
\end{align}
with
\begin{align}
  \mu_{h}^2 &= 2\lambda_0 v_h^2 ,\nno\\
  \mu_{s}^2 &= -\fr{\lambda_1 v_h^2}{2 v_s} + \fr{(3\lambda_3 +
              2\lambda_4 v_s) v_s}{6} ,\nno\\
  \mu_{hs}^2 &= (\lambda_1 + 2\lambda_2 v_s) v_h .
  \label{eq:baremass}
\end{align}
Since the off-diagonal term in the mass matrix $\mc{M}_\Phi^2$ is
non-vanishing in general, the physical Higgs states are admixtures of
$h$ and $s$.
\begin{align}
  \begin{pmatrix}
    h_1 \\ h_2
  \end{pmatrix} = \begin{pmatrix}
    \cos\theta_s & \sin\theta_s\\
    -\sin\theta_s & \cos\theta_s
  \end{pmatrix} \begin{pmatrix}
    h \\ s
  \end{pmatrix} ,
\end{align}
where the mixing angle $\theta_s$ is given by
\begin{align}
  \tan\theta_s = \fr{y}{1 + \sqrt{1 + y^2}}
\end{align}
with $y \equiv 2\mu_{hs}^2 / (\mu_h^2 - \mu_s^2)$. By diagonalizing
the mass matrix in (\ref{eq:higgs_mass_matrix}), we obtain the
tree-level Higgs boson masses as follows.
\begin{align}
  m_{h_1,\,h_2}^2 = \fr{1}{2} \l[ (\mu_h^2 + \mu_s^2) \pm (\mu_h^2 -
  \mu_s^2) \sqrt{1 + y^2} \r] .
\end{align}
We assume that $h_1$ corresponds to the SM-like Higgs boson in what follows.

The Lagrangian in (\ref{eq:lagrangian}) contains the pseudoscalar interaction in the singlet sector, which is proportional to
$\sin\theta$. After the EWSB it can be performed a chiral rotation of the singlet fermion field $\psi$ as
\begin{align}
  \psi \to \E^{\I \gamma^5 \alpha/2} \psi ,
\end{align}
and made the imaginary mass term of $\psi$ vanish by choosing
\begin{align}
  \tan\alpha = \fr{-g_S v_s \sin\theta}{m_{\psi_0} + g_S v_s
  \cos\theta} .
\end{align}
Then, the mass of the singlet fermion is given as
\begin{align}
  m_\psi &= (m_{\psi_0} + g_S v_s \cos\theta) \cos\alpha
            - g_S v_S \sin\theta \sin\alpha \nno\\
         &= \pm \sqrt{(m_{\psi_0} + g_S v_s\cos\theta)^2 + g_S^2 v_s^2
           \sin^2\theta} .
\end{align}
Note that we can always take the sign of $m_{\psi}$ to be positive by
performing the chiral rotation further.
By redefining the singlet fermion field using a chiral rotation described above, the interaction terms for the singlet fermion become
\begin{align}
  -\mc{L}^{\rm dark}_\mathrm{int} = g_S \cos\xi \,s \bar\psi \psi + g_S \sin\xi \,s
  \bar\psi \I \gamma^5 \psi ,
\end{align}
where
\begin{align}
  \cos\xi &= \fr{m_{\psi_0} \cos\theta + g_S v_s}{m_\psi} ,\nno\\
  \sin\xi &= \fr{m_{\psi_0} \sin\theta}{m_\psi} .
\end{align}

Consequently, the independent parameters for the singlet fermion
are $m_\psi$, $g_S$, and $\xi$.
The other six parameters $\lambda_0$, $\lambda_1$, $\lambda_2$, $\lambda_3$,
$\lambda_4$, and $v_s$ with $v_h\simeq 246$ GeV in the scalar sector determine
the masses $m_{h_1}$ and $m_{h_2}$, the mixing angle $\theta_s$, and self-couplings of
the two physical Higgs particles $h_1$ and $h_2$.
The cubic self-couplings $c_{ijk}$ for $h_i h_j h_k$ interactions
are given as
\begin{align}
  c_{111}
  =&~ 6 \lambda_0 v_h \cos^3 \theta_s + \l( 3 \lambda_1 + 6 \lambda_2
     v_s \r) \cos^2 \theta_s \sin\theta_s + 6 \lambda_2 v_h
     \cos\theta_s \sin^2\theta_s + (\lambda_3 + \lambda_4 v_s)
     \sin^3\theta_s , \nno\\
  c_{112}
  =& -6 \lambda_0 v_h \cos^2 \theta_s \sin\theta_s + 2 \lambda_2 v_h
     \l( 2 \cos^2\theta_s \sin\theta_s - \sin^3 \theta_s\r) \nno\\
  &  + \l(\lambda_1 + 2 \lambda_2 v_s \r) \l( \cos^3 \theta_s -
    2 \cos\theta_s \sin^2\theta_s \r) + \l( \lambda_3 + \lambda_4 v_s
    \r) \cos\theta_s \sin^2\theta_s , \nno\\
  c_{122}
  =&~ 6 \lambda_0 v_h \cos \theta_s \sin^2 \theta_s + 2 \lambda_2 v_h
     \l( \cos^3\theta_s - 2 \cos\theta_s \sin^2 \theta_s\r) \nno\\
  &  - \l(\lambda_1 + 2 \lambda_2 v_s \r) \l( 2 \cos^2 \theta_s
    \sin\theta_s - \sin^3 \theta_s \r) + \l( \lambda_3 +
    \lambda_4 v_s \r) \cos^2\theta_s \sin\theta_s , \nno\\
  c_{222}
  =& - 6 \lambda_0 v_h \sin^3 \theta_s + \l( 3 \lambda_1 + 6
     \lambda_2 v_s \r) \sin^2 \theta_s \cos\theta_s - 6 \lambda_2 v_h
     \sin\theta_s \cos^2\theta_s \nno\\
  & + (\lambda_3 + \lambda_4 v_s) \cos^3\theta_s .
\label{eq:cijk}
\end{align}
Note that $c_{112}$ is practically proportional to $\sin\theta_s$
since $\lambda_1 + 2 \lambda_2 v_s$ is vanishing if $\sin\theta_s = 0$
while the other couplings can remain non-vanishing.

\section{Galactic Center gamma-ray excess}
\label{sec:gamma_excess}

Several collaborations have analyzed the Fermi-LAT data and found
statistically significant excesses of gamma-rays at the GC over the predictions of Galactic diffuse
emission models~\cite{Goodenough,Daylan:2014rsa,Calore:2014xka}.
Consistency of the previous results was examined in Ref.~\cite{Calore:2014nla} for the intensity of the excess
at energies of 2 GeV as a function of Galactic latitude.
It was shown that those excesses typically follow the predictions of
a DM profile that is compatible with
a generalized Navarro-Frenk-White density distribution, which is
given by~\cite{Navarro:1995iw,Navarro:1996gj}
\begin{align}
\rho(r) = \rho_s \frac{(r/r_s)^{-\gamma}}{(1 + r/r_s)^{3 - \gamma}}~,
\end{align}
where $r$ is the distance from the GC.
As canonical profile we choose the scale radius $r_s  = 20$ kpc,
the slope $\gamma = 1.2$, and fix the scale density $\rho_s$ by requiring that
the local DM density $\rho=\rho_\odot = 0.4$ GeV/cm$^3$
at the location of the Solar system $r=r_\odot=8.5$ kpc.

For the present work we adopt the photon energy spectrum of the Fermi GeV excess derived by Calore, Cholis, and Weniger (CCW) in
Refs.~\cite{Calore:2014xka,Calore:2014nla} including systematic and statistical errors. The CCW spectrum rises below 1 GeV, peaking
around 2--3 GeV and has a high-energy tail up to 100 GeV. Although the excess could be explained by astrophysical sources like
millisecond pulsars~\cite{msp}, the DM annihilation still remains as an intriguing explanation.

The gamma-ray differential flux from the annihilation of a non-self-conjugate DM $\chi$
over a solid angle $\Delta \Omega$ is given by
\begin{align}
\frac{\diff N}{\diff E} =\frac{\bar J}{16\pi m_\chi^2}
\sum_{f} \langle \sigma v \rangle_f \frac{\diff N_\gamma^f}{\diff E},
\end{align}
where the sum is extended over all possible annihilation channels
into final states $f$.
Here, $\langle \sigma v \rangle_f$ is the thermally averaged annihilation cross section and $\diff N_\gamma^f / \diff E$ is the DM prompt gamma-ray spectrum per annihilation to the final state $f$. While the annihilation cross section and the spectrum per annihilation depend on particle properties,
the astrophysical factor $\bar J$ is determined from the line-of-sight (l.o.s) integral over the DM halo profile $\rho(r)$ averaged for a Region Of Interest (ROI) $\Delta\Omega$,
\begin{align}
\bar J = \frac{1}{\Delta\Omega} \int_{\Delta\Omega}
\int_\mathrm{l.o.s} \rho^2 (r(s,\,\psi)) \diff s \diff\Omega,
\end{align}
where $\psi$ is the angle from the GC.
For the ROI in the CCW analysis
($2^\circ \le \left\vert b \right\vert \le 20^\circ$ for Galactic latitude
and $\left\vert \ell \right\vert \le 20^\circ$ for Galactic longitude)
with the canonical profile for the DM halo, the value of
$\bar J$ is given as $\bar J_\mathrm{canonical} \simeq 2 \times
10^{23}$~$\mathrm{GeV}^2 / \mathrm{cm}^5$.
However, it is known that there is a significant uncertainty on the DM
density profile near the GC in particular. If the uncertainty on the
DM profile is included, the $\bar J$ value varies
from about 10\% to few times the canonical value.
In representing our analysis results we will depict the range between
0.19 and 3 times the canonical one as in Ref.~\cite{Agrawal:2014oha}.
In practice, we have extended the allowed range to [0.17, 5.3]
for numerical calculations to find the best-fit parameter point.

The expected spectra from the DM annihilations in our analysis will be
depicted alongside with the systematic uncertainties from the diagonal
elements of the covariance matrix given in \cite{Calore:2014xka,Calore:2014nla}.
However the values of goodness-of-fit $\chi^2$ are calculated
using the full covariance matrix which includes the large off-diagonal elements
due to the strong correlation of the systematic uncertainties in
different energy bins.
\texttt{LanHEP}~\cite{Semenov:2008jy} has been used to implement the SFDM
model described in Sec.~\ref{sec:model}, and the photon spectra from
the annihilation of the SFDM are obtained by using
\texttt{MicrOMEGAs}~\cite{Belanger:2014vza}.
To illustrate our analysis results we choose parameters
with the minimum $\chi^2$ while providing the DM relic density
consistent with the observed value, but by changing the scale factor
$\bar J$ explained above.

The annihilation of the SFDM through a pure scalar interaction ($\sin\xi = 0$) is velocity-suppressed. Therefore, it is inevitable to introduce
a pseudoscalar interaction in order to explain the Fermi gamma-ray
excess from the DM annihilation, taking into account its current velocity
at the GC as small as $10^{-3}$.
For the pure pseudoscalar interaction ($\sin\xi=1$),
the main annihilation channel arises from the $s$-channel pseudoscalar exchange
because the contributions from $t$ and $u$-channels and interference
terms are still $p$-wave suppressed.
Moreover, for the pure pseudoscalar interaction, the
elastic scattering of the DM on the target nuclei is
velocity-suppressed, and in consequence, the constraints from direct
detection experiments can easily be satisfied~\cite{LopezHonorez:2012kv}.
In this regard we mainly consider the pure pseudoscalar
interaction ($\sin\xi =1$), but include the analysis for the case of
mixed scalar and pseudoscalar interactions if it is necessary
to fit the gamma-ray excess avoiding other astrophysical constraints.

\begin{table}[th!]
\begin{center}
\begin{tabular}{c | c | c}
\hline\hline&&\\[-2mm]
Annihilation process & $m_\psi$ (GeV) & $m_{h_2}$ (GeV) \\[2mm]
\hline&&\\[-2mm]
$\psi \bar \psi \to h_2 \to b \bar b$ & 49.82 & 99.416 \\[2mm]
$\psi \bar \psi \to h_2 \to h_1 h_1$ & 127.5 & 213.5 \\[2mm]
$\psi \bar \psi \to h_2 \to h_2 h_2$, $h_1 h_2$ & 127.5 & 125.7 \\[2mm]
$\psi \bar \psi \to h_2 \to h_2 h_2$ & 69.2 & 35.7 \\[2mm]
\hline\hline
\end{tabular}
\end{center}
\caption{Dominant annihilation channels that can contribute to the
  gamma-ray excess are listed with the best fitted 
  masses of the DM
  and the singlet-like Higgs boson.}
\label{table:processes}
\end{table}

The detailed analysis is proceeded by finding the parameter space in each scenario 
of the main annihilation channels,
$\psi \bar \psi \to b \bar b$, $h_i h_j (\to 4b)$ for $i, \, j = 1, \, 2$.
Note that these decay modes are proven to give good fits in
model independent analyses by other groups.\footnote{For example see
  Refs.~\cite{Daylan:2014rsa,Calore:2014nla,Cline:2015qha}.}
The goodness of fits can be different in this
model dependent study partly due to theoretical and experimental bounds
that can impose constraints on the model.
On the other hand, it often occurs that several processes contribute to the DM
annihilation, depending on the mass hierarchies of particles and
couplings.
The mass values of the SFDM and the singlet-like Higgs boson $h_2$ that
turn out to give the best fits are shown in
Table~\ref{table:processes} for each dominant annihilation process,
which will be discussed in more detail in the following subsections.
We set the mass of SM-like Higgs $m_{h_1} \simeq 125$ GeV throughout our analyses
and choose a small mixing angle $\sin\theta_s \lesssim 0.12$ in order
to be compatible with the SM-like Higgs properties from the LHC
analysis results~\cite{atlascms}.

\subsection{\boldmath $\psi \bar{\psi} \rightarrow b \bar{b}$ annihilation channel}

In the model independent study in \cite{Calore:2014nla},
it was shown that the DM annihilation into $b\bar{b}$ gives
a good fit ($\chi^2 = 23.9$, $p$-value 0.35) to the gamma-ray excess data
if $m_\mathrm{DM} \simeq 48.7$~GeV and
$\langle \sigma v \rangle \simeq 1.75 \times 10^{-26} \,\mathrm{cm}^3 \,\mathrm{s}^{-1}$
for a self-conjugate DM. In this subsection we investigate
the corresponding parameter space in the SFDM model to see if such a scenario
can be realized in the model.

The SM-like Higgs boson $h_1$ would decay dominantly
to the $\psi \bar{\psi}$ pair for $m_\psi \simeq 50$~GeV
unless the coupling for the decay vertex $g_S \,\sin\theta_s$ is small.
The current analysis results on the SM-like Higgs at the LHC~\cite{atlascms}
indicate that its invisible branching ratio should be smaller than 13\% at 95\% C.L.,
provided that the production of the Higgs boson is not affected much
by unknown new physics effect.
This experimental constraint on the invisible branching ratio of
the Higgs boson implies the upper bound on the coupling $(g_S \,\sin\theta_s)^2
\lesssim 4 \times 10^{-4}$.

\begin{figure}
\begin{center}
\includegraphics[width=.6\textwidth]{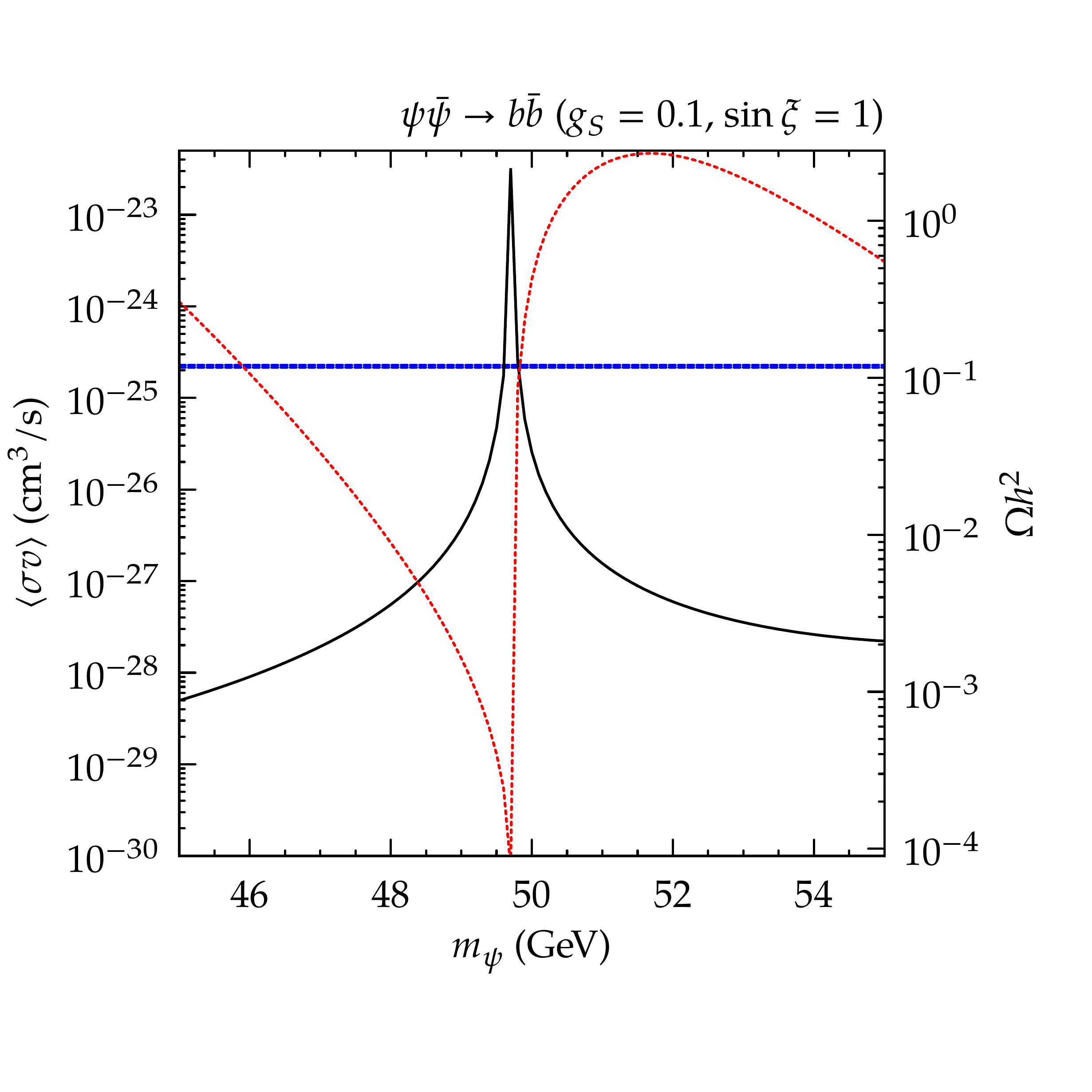}
\caption{$\langle \sigma v \rangle$ (black solid line)
and $\Omega h^2$ (red dashed line) as a function of $m_\psi$
near the resonance region ($m_\psi \sim m_{h_2}/2$) in the case of the
pure pseudoscalar interaction.
See the text for details.
For $m_\psi = 49.82$ GeV, $\Omega h^2 = 0.122$ and
$\langle \sigma v \rangle = 1.55 \times 10^{-25}$~$\mathrm{cm}^3 \,\mathrm{s}^{-1}$.}
\label{fig:bb1}
\end{center}
\end{figure}

The pair annihilation of the SFDM to a $b\bar{b}$ pair can proceed through
$s$-channel (singlet-like) Higgs exchange diagram. For the $\psi\bar\psi\rightarrow h_2
\rightarrow b\bar{b}$ process, the annihilation cross section is given by
\begin{align}
 \sigma v =
  {g_S^2~ \sin^2 2\theta_s \over 32\pi} \l({m_b\over v_h}\r)^2
  {s\over (s-m_{h_2}^2)^2+m_{h_2}^2\,\Gamma_2^2}~
  \bigg(1-{4 m_b^2 \over s}\bigg)^{3/2} N_c,
\end{align}
where $\sqrt{s}$ is the center of mass energy of the annihilation process,
$\Gamma_2$ is the decay width of $h_2$ and $N_c$ is the number of
color of the $b$ quark.
The upper bound of $(g_S\,\sin\theta_s)^2 \lesssim 4\times 10^{-4}$
from the LHC results in too small annihilation cross section to
explain the DM relic density and the Fermi gamma-ray excess, unless
there is a resonance effect with $m_{h_2} \simeq 2 m_\psi$.
Figure~\ref{fig:bb1} shows such a resonance effect on the DM relic density $\Omega h^2$
and the thermally averaged annihilation cross section $\langle \sigma v \rangle$ at the present universe.
$\Omega h^2$ (red dashed line) and $\langle \sigma v \rangle$ (black solid line) are
depicted on different $m_\psi$ values nearby the resonance region.
Here we fix the model parameters as follows: For the scalar sector,
$\lambda_0 = 0.128816$, $\lambda_1 =36.625338$~GeV, $\lambda_2 = - 0.131185$,
$\lambda_3 = - 333.447606$~GeV, $\lambda_4 = 5.648618$, and $v_s = 150.017297$~GeV,
which gives $m_{h_1} = 125.3$~GeV, $m_{h_2} = 99.416$~GeV,
$\sin\theta_s = - 0.117$. 
~\footnote{In this case $\langle \sigma v \rangle$ and $\Omega h^2$ are highly sensitive on the exact values of the parameters.}
We also set
$g_S = 0.0958$ and $\sin\xi =1$, which corresponds to the pure
pseudoscalar interaction.
As $m_\psi$ increases from 45~GeV, one can see that the relic density
of the DM $\Omega h^2$ drops down, but suddenly boosts up after
passing the resonance point $m_\psi = m_{h_2} / 2 \sim 49.6$ GeV.
For $m_\psi = 49.82$~GeV, we obtain the DM relic density $\Omega h^2 = 0.122$,
which is consistent with the current measured value from Planck~\cite{Ade:2015xua}
and the fraction of the annihilation process of $\psi \bar \psi \to b \bar{b}$ reaches 86.8\%.
However, for the same parameter values, we have the annihilation cross section
$\langle \sigma v \rangle = 1.55 \times 10^{-25} \,\mathrm{cm}^3 \,\mathrm{s}^{-1}$,
which yields too large gamma-ray flux at the GC if the canonical $\bar
J$ value is used.

\begin{figure}
\begin{center}
\includegraphics[width=.6\textwidth]{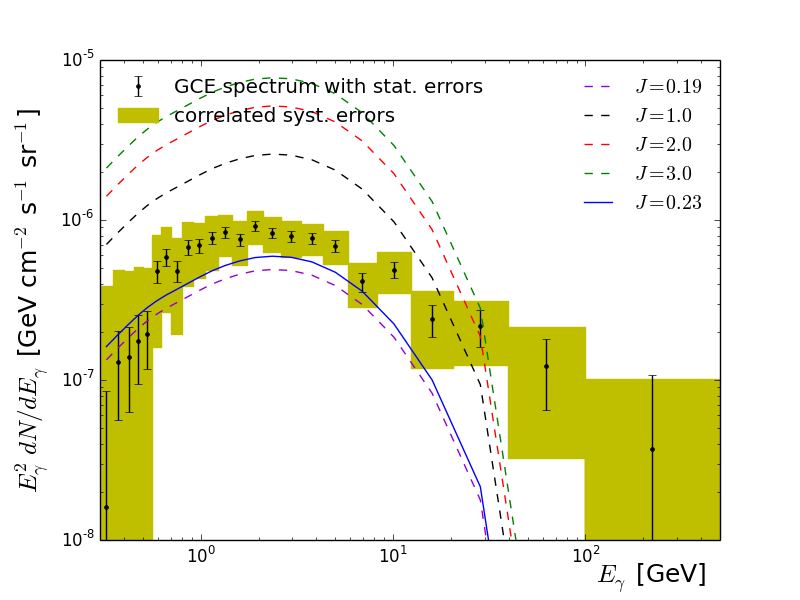}
\caption{Photon energy spectra for $m_\psi = 49.82$~GeV
with different $\cal J$ values.
The DM annihilation is dominated by $\psi \bar{\psi} \rightarrow b
\bar{b}$ process (87$\%$).
The Higgs masses are $m_{h_1} = 125.3$ GeV, $m_{h_2} = 99.416$ GeV,
and $\Omega h^2 = 0.122$, $\langle \sigma v \rangle = 1.55 \times
10^{-25}$~$\mathrm{cm}^3 \,\mathrm{s}^{-1}$.
$\chi^2 = 23.53$ ($p$-value $= 0.37$) in the best-fit parameter point
with ${\cal J} = 0.23$.}
\label{fig:bb}
\end{center}
\end{figure}

Note that the $\langle \sigma v \rangle$ value at the GC can be much larger than
the usual thermal annihilation rate at the early universe.
It is because a small difference in the center of mass energy of the
DM annihilation gives a huge difference on the annihilation
cross section in the resonance region.
As a consequence, the choice of the parameter values results in a bad fit to
the gamma-ray data, as one can see in Fig.~\ref{fig:bb}.\footnote{We have used the code and the data provided in the reference of
  \cite{Calore:2014nla} with modifications for our analysis.}
However, as mentioned above, the astrophysical $\bar J$ factor has large
uncertainties. Thus in order to have a desired gamma-ray flux at the
GC, it is required to have a smaller $\bar J$ value than the canonical
one to compensate too large $\langle \sigma v \rangle$.
We find that the best fit ($\chi^2=23.53$, $p$-value = 0.37) to the
Fermi gamma-ray excess is obtained with ${\cal J} = 0.23$ for $\bar J =
{\cal J} \times \bar{J}_\mathrm{canonical}$. The corresponding gamma-ray spectrum is shown as solid line in
Fig.~\ref{fig:bb}, where the gamma-ray spectra with other $\bar J$
values are shown as well.

Although the Fermi gamma-ray excess at the GC can be explained for a
smaller value of the $\bar J$ factor,
the annihilation cross section is too large to evade the constraints
by the observations of the gamma-ray from the dwarf spheroidal galaxies~\cite{Ackermann:2015zua}.
It sets an upper bound
$\langle \sigma v \rangle \lesssim 2 \times 10^{-26}$~$\mathrm{cm}^3 \,\mathrm{s}^{-1}$
for the non-self-conjugate DM case with $m_\textrm{DM} \sim 50$~GeV
if the majority of the annihilation products are $b \bar b$.
Furthermore, the bounds from the antiproton ratio measured by 
PAMELA~\cite{Adriani:2010rc} and AMS-02~\cite{AMS02,Giesen:2015ufa}
can strongly constrain the parameters
for $\psi\bar\psi\rightarrow b\bar{b}$ channel.
Even taking into account the uncertainties of the propagation models,
the bound should be at least
$\langle \sigma v \rangle \lesssim 2 \times 10^{-26} \,\mathrm{cm}^3
\,\mathrm{s}^{-1}$ for the non-self-conjugate DM with $m_\textrm{DM}
\sim 50$~GeV.\footnote{See \cite{Kong:2014haa} for various other
  bounds besides this.}

\begin{figure}
\begin{center}
\includegraphics[width=.6\textwidth]{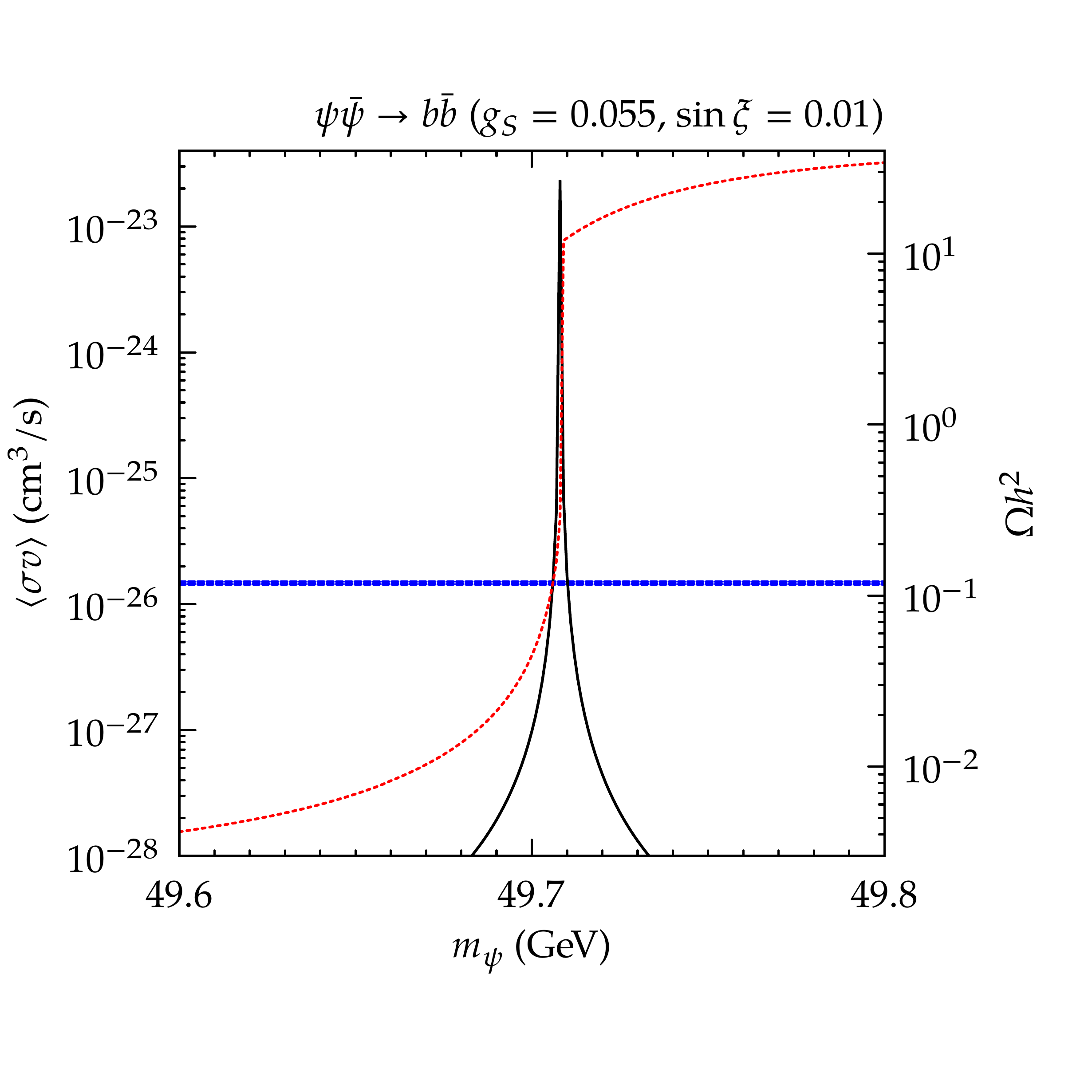}
\caption{$\langle \sigma v \rangle$ (black solid line)
and $\Omega h^2$ (red dashed line) as a function of $m_\psi$
near the resonance region ($m_\psi \sim m_{h_2}/2$) in the case of the
mixed scalar and pseudoscalar interaction.
See the text for details.
For $m_\psi = 49.706$ GeV, $\Omega h^2 = 0.118$ and
$\langle \sigma v \rangle = 1.5 \times 10^{-26}$~$\mathrm{cm}^3 \,\mathrm{s}^{-1}$.}
\label{fig:bb2}
\end{center}
\end{figure}

\begin{figure}
\begin{center}
\includegraphics[width=.6\textwidth]{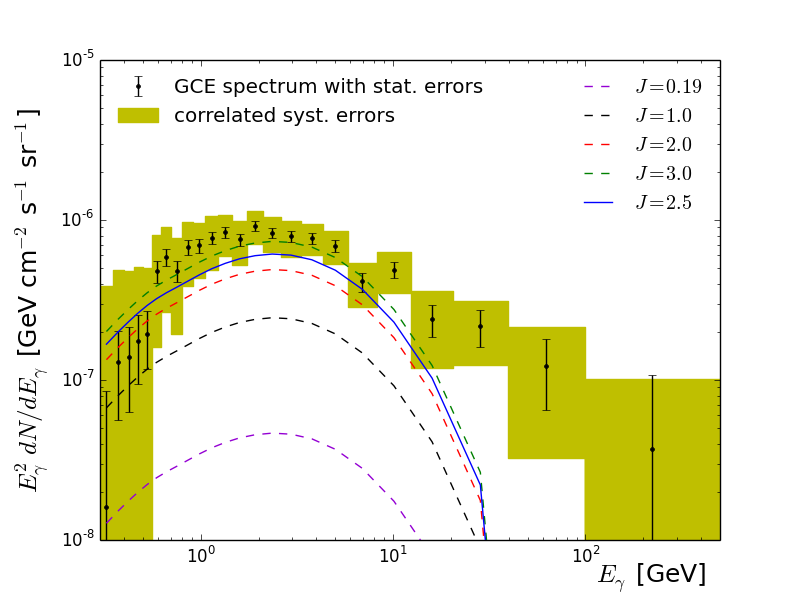}
\caption{Photon energy spectra for $m_\psi = 49.706$~GeV
with a mixture of scalar and pseudoscalar interactions
between singlet scalar and singlet fermion $\sin\xi = 0.01$.
Here $\Omega h^2 = 0.118$ and $\langle \sigma v \rangle = 1.5 \times
10^{-26}$~$\mathrm{cm}^3 \,\mathrm{s}^{-1}$.
$\chi^2 = 23.65$ ($p$-value $= 0.37$) in the best-fit parameter point
with ${\cal J} = 2.5$.}
\label{fig:bbmixed}
\end{center}
\end{figure}

To resolve this problem we can alternatively consider a mixture of scalar and pseudoscalar interactions
between singlet scalar and singlet fermion, {\em i.e.}, $\sin\xi < 1$
in order to reduce the magnitude of $\langle \sigma v \rangle$ to an acceptable level
while keeping $\Omega h^2 \sim 0.12$ and the direct detection rate of the DM small enough.
This is a viable scenario since the annihilation rate and the relic
density of the DM depend on the DM velocity in different ways
for the scalar and the pseudoscalar interactions.
For a demonstration of the effect, we set $\sin\xi = 0.01$ which is an extreme choice making the dark sector Yukawa interaction almost scalar-like 
and $g_S = 0.055$ while other model parameters unchanged. Then we obtained the annihilation
cross section and the relic density as shown in Fig.~\ref{fig:bb2}. 
The photon flux explaining the gamma-ray excess is obtained for $m_\psi = 49.706$ giving $\langle \sigma v \rangle = 1.5 \times 10^{-26}$~$\mathrm{cm}^3\,\mathrm{s}^{-1}$ and $\Omega h^2 = 0.118$. This can be seen in Fig. \ref{fig:bbmixed} and the best fit is obtained with ${\cal J} = 2.5$. The annihilation cross section is now within an acceptable range satisfying the astrophysical constraints mentioned above. 
In addition the spin independent cross section of the DM recoiling against neutron or proton is still around $6.3 \times 10^{-48}~{\rm cm}^2$ which is below the bounds from various direct detection experiments. This is because of the small mixing angle $\theta_s$ although we considered mostly scalar-like interaction $s \bar \psi \psi$.

\subsection{\boldmath $\psi \bar{\psi} \rightarrow h_1 h_1$ annihilation channel}

\begin{figure}[tb!]
  \begin{center}
    \includegraphics[width=0.32\textwidth]{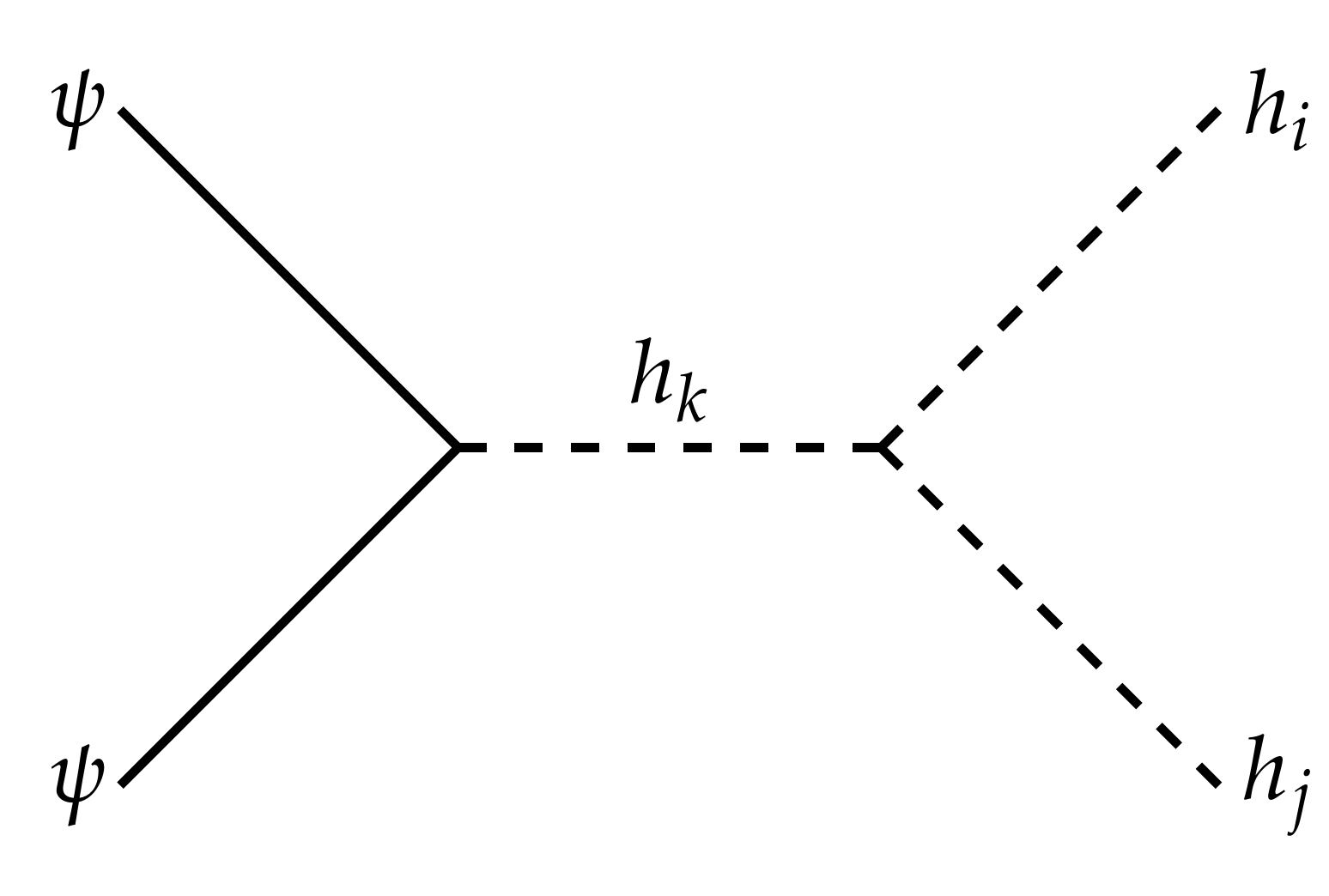}
    \includegraphics[width=0.32\textwidth]{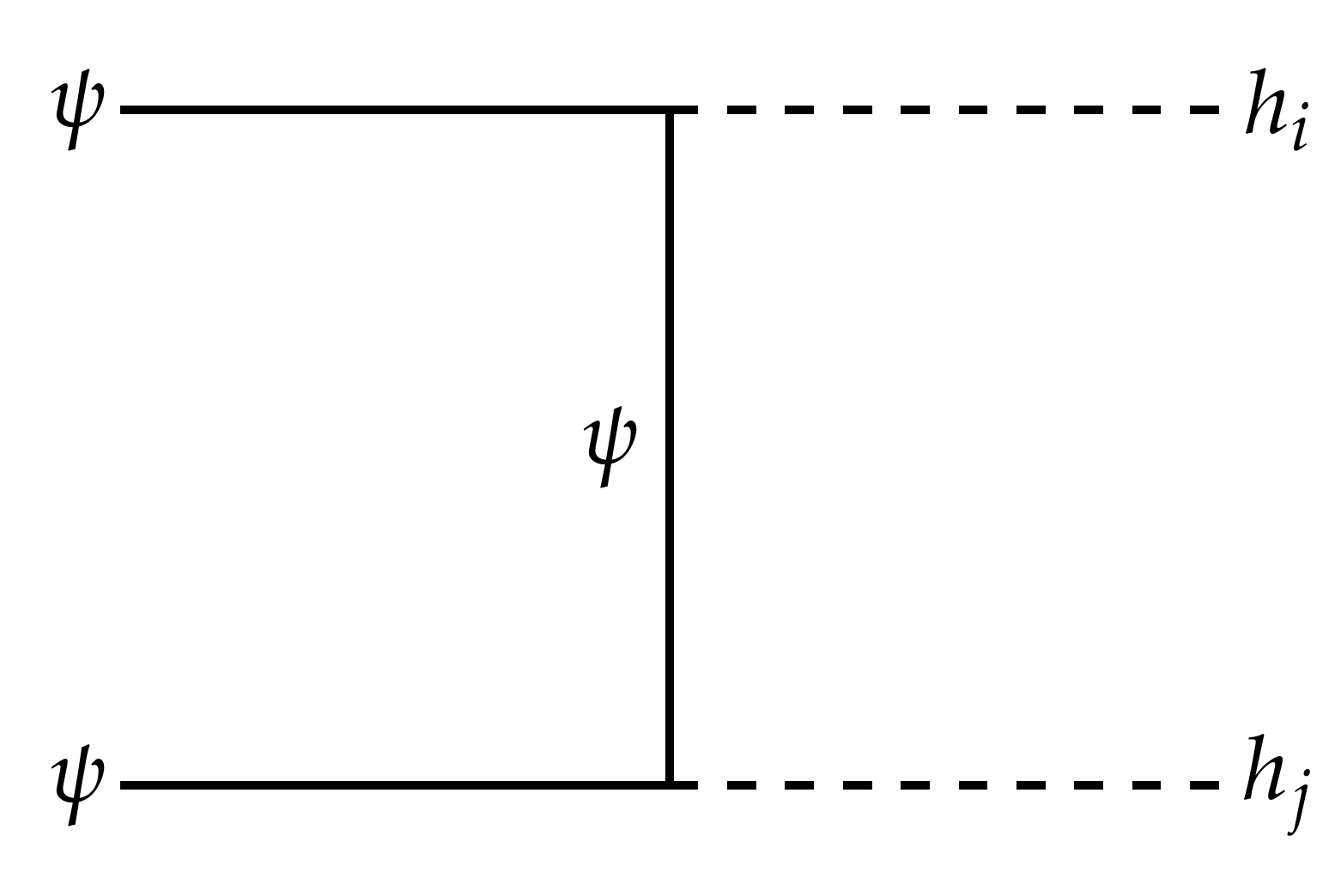}
    \includegraphics[width=0.32\textwidth]{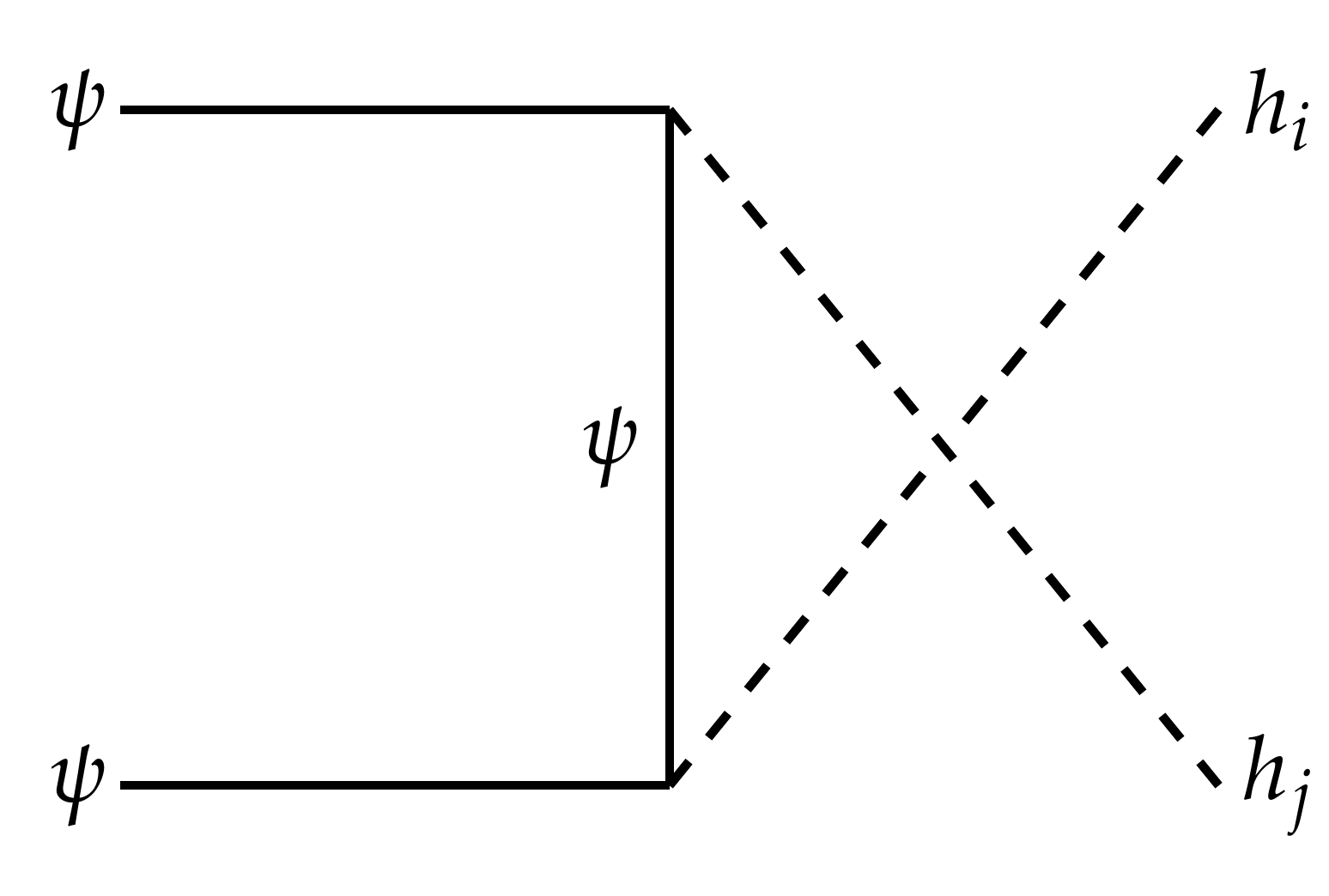}
  \end{center}
  \caption{Diagrams for $\psi\psi \to h_i h_j$ annihilation
    processes. $i$, $j$, $k = 1$, 2.}
  \label{fig:diagram}
\end{figure}

The annihilation into a non-relativistic pair of the Higgs boson can
give a good fit to the gamma-ray
excess~\cite{Agrawal:2014oha,Calore:2014nla}.
It has been shown in Ref.~\cite{Calore:2014nla} that the best $\chi^2 =
29.5$ ($p$-value $= 0.13$) is obtained with $m_\psi \simeq m_{h_1}
\simeq 125.7$~GeV and $\langle \sigma v \rangle = 5.33 \times 10^{-26}$~$\mathrm{cm}^3
\,\mathrm{s}^{-1}$ for a self-conjugate DM that annihilates into $h_1
h_1$.
Here we investigate if this scenario can be realized in the SFDM
model. Diagrams for the annihilation processes $\psi\bar\psi
\rightarrow h_i h_j$ are shown in Fig.~\ref{fig:diagram}.

For the pure scalar interaction, {\em i.e.}, $\sin\xi= 0$, the
annihilation cross section vanishes in the zero-velocity limit.
On the other hand, for the pure pseudoscalar interaction
{\em i.e.}, $\sin\xi = 1$, only $s$-channel diagram contributes to the
annihilation cross section in the zero-velocity limit.
Therefore, magnitudes of cubic couplings $c_{ijk}$ of the Higgses given in
(\ref{eq:cijk}) play important roles for the processes 
$\psi\bar\psi
\rightarrow h_i h_j$ in the zero-velocity limit.
The annihilation cross section for $\psi\bar\psi \rightarrow h_2
\rightarrow h_1 h_1$, which would provide the most important
contribution to the $h_1 h_1$ channel, is given by
\begin{align}
 \sigma v
 = \fr{g_S^2}{32\pi} \sqrt{1-\fr{4m_{h_1}^2}{s}} \,
   \fr{c_{112}^2 \cos^2 \theta_s}{(s - m_{h_2}^2)^2 + m_{h_2}^2 \Gamma_{2}^2}.
\end{align}

If $\psi\bar\psi$ annihilation into $h_1 h_1$ channel opens,
the annihilation rates into $WW$ and $ZZ$ channels can be sizable as well.
The channels with SM gauge bosons are known to yield relatively a bad fit to
the Fermi gamma-ray excess~\cite{Calore:2014nla}.
Thus, good fits can be obtained if the DM annihilation rate
into $h_1 h_1$ is dominant over $WW$ and $ZZ$ by having a large value
of the cubic coupling $c_{112}$ for given masses $m_{h_1}$, $m_{h_2}$,
and the mixing angle $\theta_s$.

\begin{figure}
\begin{center}
\includegraphics[width=.6\textwidth]{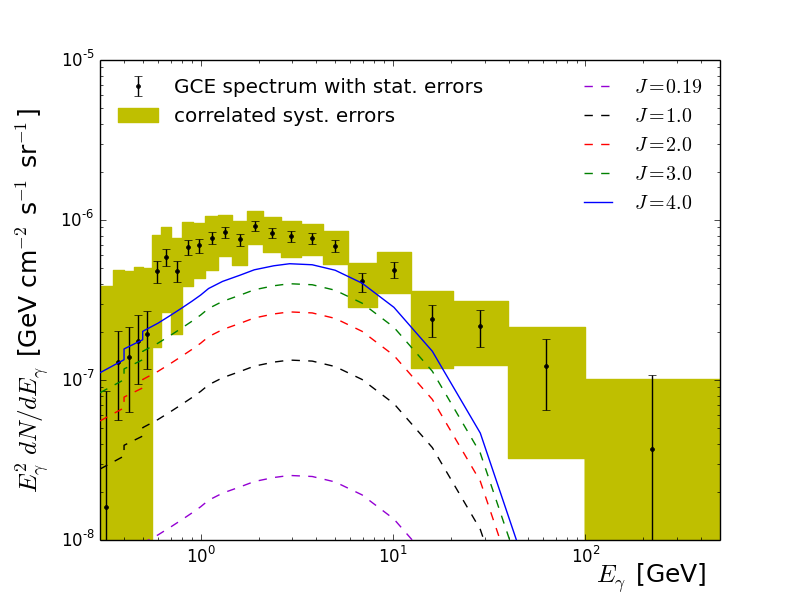}
\caption{Photon energy spectra for $m_\psi = 127.5$~GeV with
different $\mathcal{J}$ values.
The DM annihilation is dominated by $\psi \bar{\psi} \to h_1 h_1$
process (96\%).
The Higgs masses are $m_{h_1} = 124.9$~GeV, $m_{h_2} = 213.5$~GeV,
and $\Omega h^2 = 0.12$, $\langle \sigma v \rangle = 2.11 \times
10^{-26}$~$\mathrm{cm}^3 \,\mathrm{s}^{-1}$.
$\chi^2 = 31.32$ ($p$-value $= 0.09$) in the best-fit parameter point
with ${\cal J} = 4.0$.}
\label{fig:h1h1}
\end{center}
\end{figure}

As a specific example, we choose the parameter values
as follows: $m_\psi = 127.5$ GeV, $g_S = 0.098$, and $\sin\xi=1$ for
interactions of the SFDM, and $\lambda_0 =0.1315$, $\lambda_1 = 1237.8$~GeV,
$\lambda_2 = - 2.0$, $\lambda_3 = - 820.5$~GeV, $\lambda_4 = 9.39$,
and $v_s = 306.15$~GeV for the scalar sector, which yield Higgs masses
$m_{h_1}=124.9$~GeV, $m_{h_2}=213.5$~GeV, and the mixing angle $\sin\theta_s=-0.11$
with the cubic couplings $c_{111}=149.0$~GeV and $c_{112}=268.8$~GeV.
With this parameter choice, $\psi\bar\psi \rightarrow h_1 h_1$
becomes the most dominant annihilation process ($\simeq 96\%$ for the
annihilation at the GC), and we have $\Omega h^2 = 0.12$ and
$\langle \sigma v \rangle = 2.11 \times 10^{-26}$~$\mathrm{cm}^3 \,\mathrm{s}^{-1}$.
The $\langle \sigma v \rangle$ value is rather smaller than desired, so a large $\mathcal J$ factor
is necessary to explain the Fermi gamma-ray excess.
For $\mathcal{J}=4$, we obtain the best fit ($\chi^2 = 31.3$, $p$-value $= 0.09$)
and the corresponding gamma-ray spectrum is shown in
Fig.~\ref{fig:h1h1}.

The important astrophysical bounds to be considered in this scenario are the observations of gamma-rays from the dwarf spheroidal galaxies and the antiproton ratio.
The annihilation cross section value that we obtained is below the upper bound from the
dwarf spheroidal galaxies for a non-self-conjugate DM, $5.2 \times
10^{-26}$~$\mathrm{cm}^3 \,\mathrm{s}^{-1}$ around $m_\psi \simeq
125$~GeV, assuming that the dominant annihilation process is $\psi \bar \psi \to
b \bar b$~\cite{Ackermann:2015zua}.
For the four-body final states, {\em i.e.}, $\psi \bar \psi \to
b \bar b b \bar b$, the authors of Ref.~\cite{Dutta:2015ysa} extracted
rough bounds, but they tend to be less constrained than the two-body
case. Therefore, the best-fit parameter in our analysis is safe from
the gamma-ray bound from the dwarf spheroidal galaxies.
On the other hand, the $4b$ final state has been included in the
analysis of \cite{Cline:2015qha} in light of the search results on the
antiproton excess combined from BESS~\cite{BESS}, CAPRICE~\cite{Boezio:2001ac}, and PAMELA~\cite{Adriani:2010rc}.
The value of the annihilation cross section for
$m_\psi \simeq 125$~GeV that we obtained is below this bound if the uncertainties of the propagation models are included.

\subsection{\boldmath $\psi \bar\psi \to h_1 h_1$/$h_1 h_2$/$h_2 h_2$ annihilations
in the mass-degenerate case}

Another interesting possibility that is closely related to the $h_1 h_1$ channel in
the previous subsection arises if two Higgses $h_1$ and $h_2$ are almost degenerate in mass.
Then all the annihilation modes $\psi \bar \psi \to h_{1,2} \to h_1
h_1$/$h_1 h_2$/$h_2 h_2$ have no differences in the phase space and
provide the same spectral shape for the photon energy spectrum.

We find the parameter choice giving one of the best fit for the galactic
gamma-ray excess at $m_\psi =127.5$~GeV
is $\lambda_0 = 0.13$, $\lambda_1 = 112.49$~GeV, $\lambda_2 = -0.20$,
$\lambda_3 = -898.97$~GeV, $\lambda_4 = 5.97$, $v_s = 277.01$~GeV,
and $g_S = 0.085$, which gives $m_{h_1} = 125.5$~GeV, $m_{h_2} = 125.7$~GeV,
$\sin\theta_s = -0.11$, $c_{222} = 705.8$~GeV, and $c_{122} = -177.8$~GeV.
In this case the process $\psi \bar \psi \to h_2 \to h_2 h_2$ is the
most dominant since the amplitude is not suppressed by the smallness of the mixing angle
$\sin\theta_s$ and the magnitude of $c_{222}$ is much larger than that of $c_{122}$. Note that the values of $c_{222}$ (and $c_{122}$) can be arbitrarily given without affecting the scalar masses and mixing angle. 
The annihilation cross section for $\psi\bar\psi \rightarrow h_2
\rightarrow h_2 h_2$ is given by
\begin{align}
 \sigma v
 = \fr{g_S^2}{32\pi} \sqrt{1-\fr{4m_{h_2}^2}{s}} \,
   \fr{c_{222}^2 \cos^2 \theta_s}{(s - m_{h_2}^2)^2 + m_{h_2}^2
  \Gamma_{2}^2}.
  \label{eq:ann_h2h2}
\end{align}

\begin{figure}
\begin{center}
\includegraphics[width=.6\textwidth]{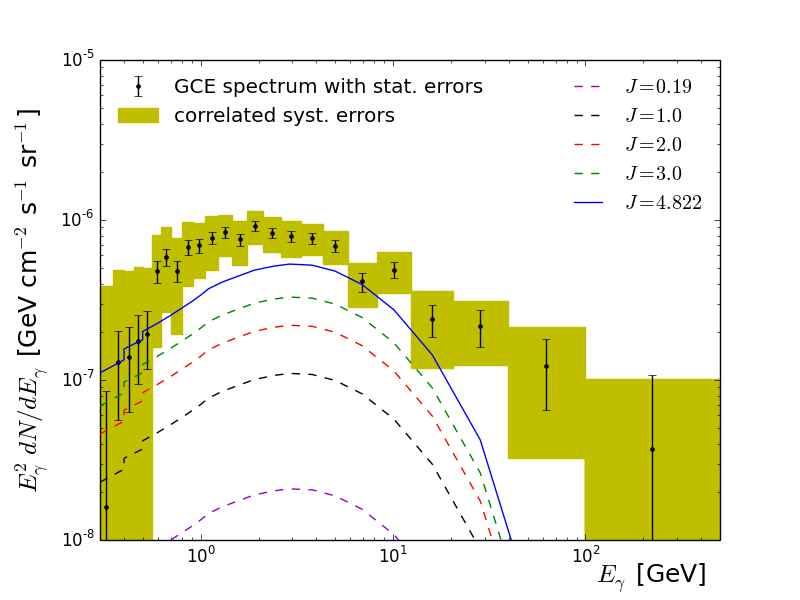}
\caption{Photon energy spectra for $m_\psi = 127.5$~GeV with
different $\mathcal{J}$ values.
The DM annihilation is dominated by $\psi \bar{\psi} \to h_i h_j$ $(i, \, j = 1,\, 2)$
process ($\simeq 100\%$).
The Higgs masses are $m_{h_1} = 125.5$~GeV, $m_{h_2} = 125.7$~GeV,
and $\Omega h^2 = 0.12$, $\langle \sigma v \rangle = 1.71 \times
10^{-26}$~$\mathrm{cm}^3 \,\mathrm{s}^{-1}$.
$\chi^2 = 30.8$ ($p$-value $= 0.1$) in the best-fit parameter point
with ${\cal J} = 4.822$.}
\label{fig:h1h2deg}
\end{center}
\end{figure}

With those parameters we obtain the DM relic density $\Omega h^2 =
0.12$ and the total annihilation cross section $\langle \sigma v \rangle = 1.71 \times
10^{-26}$~$\mathrm{cm}^3 \,\mathrm{s}^{-1}$.
The fraction of the DM annihilation rate to $h_2 h_2$ is 88.2\%
while that to $h_1 h_2$ is 11.6\%.
The annihilation cross section value is in the allowed region for
the constraints from the dwarf spheroidal galaxies~\cite{Dutta:2015ysa} and
also from the antiproton measurements~\cite{Cline:2015qha}, but
$\mathcal{J} = 4.822$ is required to explain the Fermi gamma-ray excess.
We obtain an acceptable fit ($\chi^2 = 30.8$, $p$-value = 0.1) with
this large $\mathcal J$ factor. The corresponding gamma-ray spectrum is shown in
Fig.~\ref{fig:h1h2deg}.

\subsection{\boldmath $\psi \bar{\psi} \rightarrow h_2 h_2$ annihilation channel}

One of interesting features of the SFDM model is that there is a mediator $h_2$, which
is an important new particle for realizing what is called the secluded WIMP scenario
in our setup~\cite{pospelov:2008,Kim:2009ke}.
Various mediator particles have been introduced in several
model independent studies to explain the Fermi gamma-ray excess
from the production of a pair of the light mediator particle with
its subsequent cascade decay into SM
fermions~\cite{Abdullah:2014lla,Berlin:2014pya,Cline:2015qha,Rajaraman:2015xka,
Dutta:2015ysa,Fortes:2015qka,Elor:2015tva}.
In the model independent study of Ref.~\cite{Dutta:2015ysa},
it was shown that the DM annihilation into a pair of new particles ($\phi\phi$)
with subsequent $\phi$ decay to $b\bar{b}$, gives a good fit
($\chi^2 = 23.1$) if $m_\mathrm{DM} = 65$ GeV, $m_\phi=m_\mathrm{DM}/2$ and
$\langle \sigma v \rangle = 2.45 \times 10^{-26}$~$\mathrm{cm}^3 \,\mathrm{s}^{-1}$
for a self-conjugate DM.
In this subsection we consider the corresponding channel in our model and find the best-fit parameters by varying the masses.

\begin{figure}
\begin{center}
\includegraphics[width=.6\textwidth]{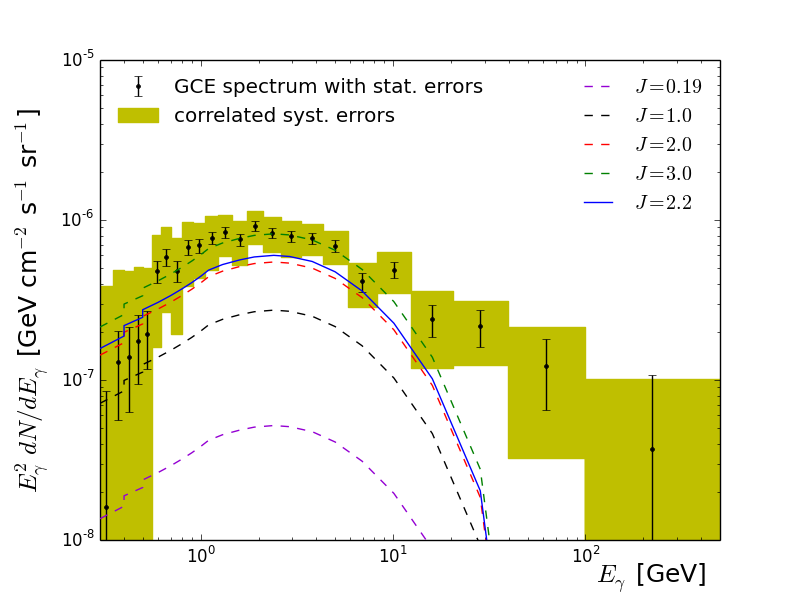}
\caption{Photon energy spectra for $m_\psi = 69.2$~GeV with
different $\mathcal{J}$ values.
The DM annihilation is dominated by $\psi \bar{\psi} \to h_2 h_2$
process ($\simeq 100$\%).
The Higgs masses are $m_{h_1} = 125.1$~GeV, $m_{h_2} = 35.7$~GeV,
and $\Omega h^2 = 0.121$, $\langle \sigma v \rangle = 2.26 \times
10^{-26}$~$\mathrm{cm}^3 \,\mathrm{s}^{-1}$.
$\chi^2 = 23.19$ ($p$-value $= 0.39$) in the best-fit parameter point
with ${\cal J} = 2.2$.}
\label{fig:h2h2}
\end{center}
\end{figure}

The fraction of the DM annihilation rate for $\psi \bar \psi \to h_2 h_2$
with $m_\psi \simeq 70$~GeV and $m_{h_2} < m_\psi$ can easily become as
large as 100\% by taking suitable parameter values of the model.
We also found that the best-fit spectrum is obtained
if $m_{h_2} \sim m_\psi / 2$, as pointed out in the model independent
study~\cite{Dutta:2015ysa}. Finding parameters for the good fits, we further consider the bound from the search of exotic Higgs decays \cite{atlascms} due to the decay mode $h_1 \to h_2 h_2 \to 4b$.
Our choice of model parameters for the best fit is as follows:
$m_\psi = 69.2$~GeV, $g_S=0.056$, $m_{h_1}=125.1$~GeV, $m_{h_2} = 35.7$~GeV,
$\sin\theta_s = 0.025$, and $c_{222} = 215.1$~GeV
from $\lambda_0 = 0.13$, $\lambda_1 = 4.5$~GeV, $\lambda_2 = -0.0055$,
$\lambda_3 = -391.51$~GeV, $\lambda_4 = 2.20$, and $v_s = 276.21$~GeV.
With these parameters the dominant contribution for the DM annihilation comes from
the $\psi\bar\psi \rightarrow h_2 \rightarrow h_2 h_2$ process.
See Eq.~(\ref{eq:ann_h2h2}) for the corresponding annihilation cross section formula.
Here the relic density $\Omega h^2 = 0.121$ and
$\langle \sigma v \rangle = 2.26 \times 10^{-26}$~$\mathrm{cm}^3\,s^{-1}$.
We find that a good value of $\chi^2 = 23.19$ ($p$-value $= 0.39$) can
be obtained with a moderate $J$ factor value, $\mathcal{J}=2.2$.
The corresponding gamma-ray spectrum is shown in Fig.~\ref{fig:h2h2}.

The astrophysical bounds can be important like the previous scenarios.
The analysis results on the gamma-ray search coming from the dwarf
spheroidal galaxies in \cite{Dutta:2015ysa} show that the upper bound of
the annihilation rate of $\psi\bar\psi \rightarrow h_2 h_2 \rightarrow 4 b$
is expected to be at least $3.3 \times 10^{-26}~\mathrm{cm}^3\,\mathrm{s}^{-1}$
with $m_\psi =70$ GeV for a non-self-conjugate DM. Therefore, our
$\langle \sigma v \rangle$ value from the best-fit parameters is below the current upper bound.
Following the antiproton bound for the $4b$ final-state analyzed in
\cite{Cline:2015qha}, as commented in previous subsections,
$\langle \sigma v \rangle = 2.26 \times 10^{-26}$~$\mathrm{cm}^3\,\mathrm{s}^{-1}$ at
$m_\psi \simeq 70$ GeV is below the bound if the
uncertainties in the propagation models are taken into consideration.

\section{Conclusions}
\label{sec:concl}

\noindent
In this paper we considered a model with the SFDM,
in which the mechanism of the thermal freeze-out is secluded from
its observations in the direct detection and collider experiments.
In addition to suppressing the mixing angle between the SM Higgs boson
and the singlet scalar we introduced a pseudoscalar interaction at the
singlet sector to amplify the secludedness.
In this type of model the DM search is generically difficult in the
direct detection and collider experiments due to the secludedness.
Nonetheless, various indirect detection results can shed light
on the parameter space to be probed.
As an observational guide, we applied the model to the recent results
on a few GeV level gamma-ray excess at the GC revealed by the analyses on
the Fermi-LAT data, which has been a hot issue in both theoretical and
experimental sides to date.

As a concrete analysis we adopted the results by CCW and applied the
systematic uncertainties estimated by them.
Then we categorized the annihilation processes depending on the final states,
$\psi \bar \psi \to b \bar{b}$, $h_1 h_1$, $h_1 h_1$/$h_1 h_2$/$h_2
h_2$, or $h_2 h_2$, where the latter three channels are cascade
processes producing multiple SM fermions or gauge bosons.
The direction of our paper is not just explaining the gamma-ray excess
but finding the model parameter values preferred by the observation
and the constraints for the future study of the SFDM model.
In this regard, other astrophysical constraints
such as gamma-ray bounds from the dwarf spheroidal galaxies (by Fermi-LAT)
and searches of antiproton excesses (by PAMELA and AMS-02) 
together with LHC bounds on the Higgs boson are taken into account in our study.
In order to satisfy these bounds we keep the value of the observed relic
density of the DM while adopting the large uncertainties of the DM
density profile near the GC to obtain the best-fit parameter point in
each channel.
Our analysis found that the excess can be obtained with the similar level of 
$\chi^2$ values as those in various model independent searches,
particularly for $\psi \bar \psi \to h_2 \to b \bar{b}$ and $h_2 h_2$
channels with $(m_\psi, \, m_{h_2}) = (49.82~\mathrm{GeV}, \,
99.416~\mathrm{GeV})$, $(69.2~\mathrm{GeV}, \, 35.7~\mathrm{GeV})$,
respectively considering the pure pseudoscalar interaction in the dark sector. However the former case is again strongly constrained from astrophysical and collider bounds commented above so a mixture of the singlet and pseudoscalar interaction in the dark sector is needed.

Although it is not easy to find the signals of the secluded SFDM model at the current level of the LHC, we may observe those in the future high luminosity LHC or next generation colliders.
In particular various channels by trilinear Higgs interactions can provide interesting signals. We will proceed the collider analyses for the parameter space found in this work, targeting their signatures at the LHC and future colliders~\cite{future}.

\section*{Acknowledgements}
YGK is supported by the Basic Science Research Program through the National Research
Foundation of Korea (NRF) funded by the Korean Ministry of Education, Science and
Technology (NRF-2013R1A1A2012392). KYL is supported by Basic Science Research Program
through the National Research Foundation of Korea (NRF) funded by the Ministry of Science, ICT and Future Planning (Grant No. NRF-2015R1A2A2A01004532). SS thanks Jong-Chul~Park for useful discussions on various astrophysical constraints. SS also thanks the University of Tokyo and the Korea Institute for Advanced Study for hospitality and support during the completion of this work.

\end{document}